\documentclass[usenatbib,aas_macros]{mn2e}

\usepackage{graphicx}
\usepackage{amssymb}

\title[INTEGRAL SMC]{INTEGRAL observations of the Small Magellanic Cloud}

\author[V.A. McBride et al.]{V.A. ~McBride$^{1}$, M.J. ~Coe$^{1}$,
A.J. ~Bird$^{1}$, A.J. Dean$^{1}$, A.B. ~Hill$^{1}$, \and
K.E.~McGowan$^{1}$, M.P.E. Schurch$^{1}$, A. ~Udalski$^{2}$,
I. Soszynski$^{2}$, \and M. ~Finger$^{3}$, C.A. ~Wilson$^{3}$,
R.H.D. Corbet$^{4}$ and I. Negueruela$^{5}$\\
$^{1}$ School of Physics and Astronomy, Southampton University, SO17 
1BJ, UK\\
$^{2}$ Warsaw University Observatory, Aleje Ujazdowskie 4, 00-478 Warsaw, Poland\\
$^{3}$ NASA/MSFC, Huntsville, AL 35812, USA\\
$^{4}$ USRA, Mail Code 662, NASA/GSFC, Greenbelt, MD 20771, USA\\
$^{5}$ Departamento de Fisica, Ingeniera de Sistemas y Teora de la Seal, 
Escuela Politcnica Superior, \\
University of Alicante, Ap.99, 03080 Alicante, Spain.}

\begin{document}

\date{accepted version : 3 Sep 2007}

\pagerange{\pageref{firstpage}--\pageref{lastpage}} \pubyear{2002}

\maketitle

\label{firstpage}

\begin{abstract}

The first INTEGRAL observations of the Small Magellanic Cloud
(carried out in 2003) are reported in which two sources are clearly
detected.  The first source, \hbox{SMC~X-1}, shows a hard X-ray eclipse and
measurements of its pulse period indicate a continuation of the long-term spin-up now covering $\sim$30
years. The second source is likely to be a high mass X-ray binary, and shows a potential periodicity of 6.8\,s in the IBIS lightcurve.  An exact X-ray or optical counterpart cannot be designated, but a number of proposed counterparts are discussed.  One of these possible counterparts shows  a
strong coherent optical modulation at $\sim2.7$\,d, which, together with the measured hard X-ray pulse period, would lead to this INTEGRAL source being classified as the fourth known high mass Roche lobe
overflow system.  

\end{abstract}


\section{Introduction and background}

X-ray satellite observations have revealed that the Small Magellanic
Cloud (SMC) contains an unexpectedly large number of High Mass X-ray
Binaries (HMXB). At the time of writing, $\sim$60 known or probable
sources of this type have been identified in the SMC and they continue
to be discovered at a rate of about 2-3 per year, although only a
small fraction of these are active at any one time because of their
transient nature.  Unusually (compared to the Milky Way and the LMC)
all the X-ray binaries so far discovered in the SMC are HMXBs, and
equally strangely, only one of the objects is a supergiant system
(\hbox{SMC~X-1}), all the rest are Be/X-ray binaries. A review of
these systems may be found in Haberl \& Sasaki (2000) and Coe et
al. (2005).

In this paper observations by the INTEGRAL observatory are reported
which primarily cover the hard X-ray regime. Since the fully coded
field of view of IBIS (Imager on Board INTEGRAL Satellite) extends
over 9 degrees, it is possible to view the entire SMC at
once. Detailed timing and imaging results are presented.

\section{INTEGRAL Observations}

The SMC was observed with INTEGRAL during
2003 July and August, corresponding to satellite revolutions 94 and
95, or 155\,ks of observing time. 
Standard processing of the IBIS and JEM-X (Joint European Monitor for X-rays) data was performed using
INTEGRAL Off-line Science Analysis (OSA Goldwurm et al 2003) software version 5.1.  After
basic data correction, good time handling, dead time correction,
background correction and event binning, sky images were created for
each individual pointing ($\sim2000$\,s) in the following energy
bands: 20--40, 40--60 and 60--100\,keV with IBIS and 3--10 and
10--35\,keV with JEM-X.

\begin{figure}
\begin{center}
\includegraphics[width=60mm,angle=0]{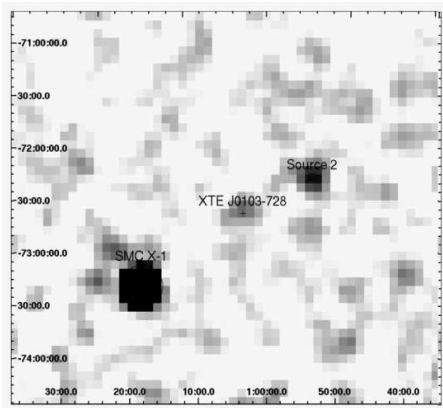}
\caption{IBIS 20--40 keV image of the SMC. The position of XTE J0103-728 = SXP6.85 is also shown (Haberl et al. 2007).}
\label{fig:xtej2a}
\end{center}
\end{figure}

To remove systematic effects in the ISGRI (INTEGRAL Soft Gamma-Ray Imager) detector
plane, successive science windows are dithered, resulting in slightly
different pointing positions.  Thus images incorporating more than one
science window need to be mosaicked together.  In addition this mosaicking improves the signal and allows one to search for fainter sources which may not appear in a single science window.  In the IBIS 20--40\,keV
mosaic, a section of which is shown in Figure~\ref{fig:xtej2a}, two
sources were apparent above the detection limit:

\begin{enumerate}  

\item SMC~X-1, with a detection significance of $100\sigma$ and an
average flux of
$F_{20-40\,\rm{keV}}=3.2\times10^{-10}$\,erg\,cm$^{-2}$\,s$^{-1}$,
which corresponds to a luminosity of $1.4\times10^{38}$\,erg\,s$^{-1}$
at a distance of 60\,kpc (Dolphin et al 2001).  The source position is
determined as $01^{\rm h}17^{\rm m} 09^{\rm s},-73^\circ 26'47''$ with
a 90\% error circle of radius $24''$.

\item Source 2, with a detection significance of $9\sigma$ and an average
flux of
$F_{20-40\,\rm{keV}}=2.6\times10^{-11}$\,erg\,cm$^{-2}$\,s$^{-1}$,
which corresponds to a luminosity of $1.1\times10^{37}$\,erg\,s$^{-1}$
at a distance of 60\,kpc (Dolphin et al 2001).
The source position is determined as $00^{\rm h} 54^{\rm m} 10^{\rm s}, -72^\circ 25' 44 ''$ 
with a 90\% error circle of radius $3.4'$.

\end{enumerate}

Although SMC~X-1 was apparent in JEM-X observations, Source~2 was not
detected with JEM-X.

In Figure~\ref{fig:xtej2a} we also
overplot the position of source \hbox{XTE~J0103-728} (a 6.8s X-ray pulsar),
as determined from XMM observations (Haberl et al. 2007).  Please
note that although this source is not detected in this INTEGRAL
observation, the source position is plotted on Figure~\ref{fig:xtej2a}
to aid the discussion in Sect.~\ref{SectDiscussion} that Source~2 and
\hbox{XTE~J0103$-$728} are two different sources.

\section{Timing analysis}

\subsection{SMC X--1}

\begin{figure}
\begin{center}
\includegraphics[width=90mm,angle=-0]{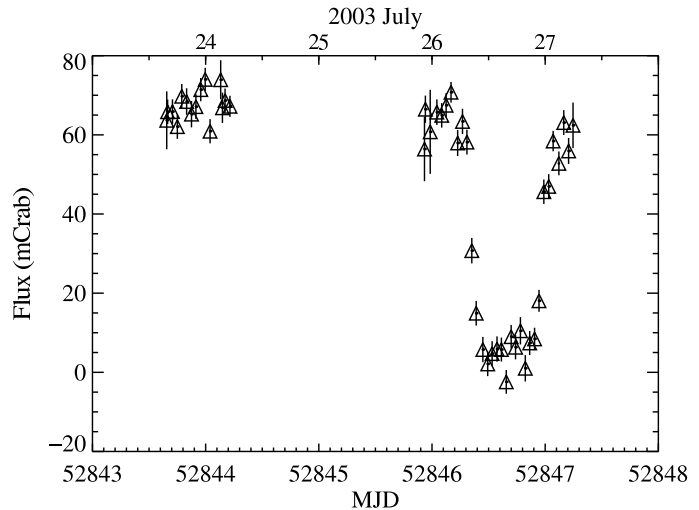}
\caption{Flux v time in the 20-40 keV band for SMC X-1.}
\label{fig:lc1}
\end{center}
\end{figure}

A lightcurve at science window time resolution ($\sim2000$\,s) was
extracted in the 20--40 \,keV energy range for the epoch MJD 52843 to
MJD 52848.  Figure~\ref{fig:lc1} clearly shows the source undergoing
a hard X-ray eclipse.

SMC~X-1 has shown a constant spin-up throughout the history of the
source.  The spin period of SMC~X-1, by far the brightest source in
the mosaic, was investigated by extracting time and energy tagged
events from the ISGRI detector for 2.95 hours (3 Science Windows)
starting at MJD 52843.96.  These science windows were chosen as they
were consecutive in time and in these observations SMC~X-1 was very
close (within $3^{\circ}$) to the telescope pointing axis.  Events
were filtered by selecting only those which have a Pixel Illumination
Fraction (PIF) of 1.  The PIF is a number, relevant to coded-mask
telescopes, between 0 and 1 designating
the fraction of a pixel that is illuminated by a source at a given
position on the sky.  Choosing events based on the criterion that PIF
= 1 increases the signal to noise by ensuring that only events from
pixels which are totally illuminated by the source contribute.
However, photons from other sources in the IBIS field of view may also
contribute to events in these pixels as background, and hence, any
methods (like this one) which do not involve the construction of a
shadowgram are subject to contamination by all other sources in the
field.  In general, even for an event list selected on the criterion
that the PIF for a specific source should equal 1, 95\% of the events
will be background.

A further filtering, allowing only events at energies between 20 and
100\,keV to contribute to the signal, was applied to the event list.
The event times were corrected to the Solar System barycentre and
corrected to the barycentre of the neutron star-supergiant system,
using the orbital ephemeris from Wojdowski et al. (1998).

\begin{figure}
\begin{center}
\includegraphics[width=80mm,angle=-0]{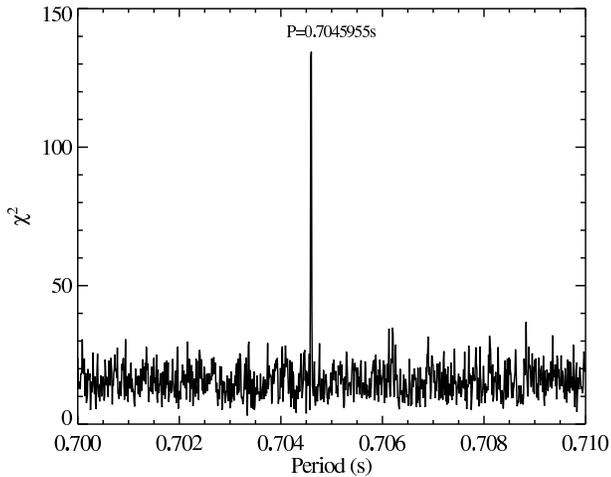}
\caption{Results of the epoch-folding search for SMC X-1}
\label{fig:pdg}
\end{center}
\end{figure}

\begin{figure}
\begin{center}
\includegraphics[width=80mm,angle=-0]{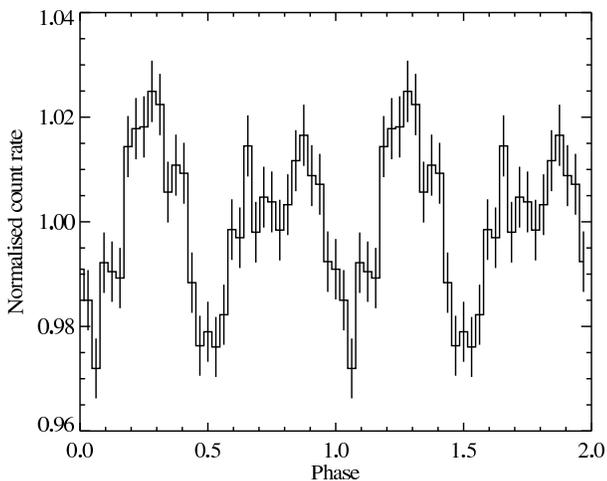}
\caption{Pulse profile of SMC X--1 in the 20--100 keV band}
\label{fig:flc}
\end{center}
\end{figure}

According to the pulse ephemeris in Naik \& Paul (2004), the source
is expected to be spinning at a period of 0.7052\,s at the start of
the observations in our study.  An epoch folding technique (Leahy et al., 1983) was used to determine
the $\chi^2$ statistic for a number of test periods in the range 0.70 to
0.71\,s at a resolution of 0.0001\,s.  The pulsation
period is taken as that with the greatest deviation from a constant
flux. Figure~\ref{fig:pdg} shows the $\chi^2$ statistic plotted
against period for the SMC~X-1 data.  The detection of pulsations at
0.704596$\pm$0.000008\,s is clearly visible.  Figure~\ref{fig:flc}
represents the 20--100\,keV pulse profile at this pulse period. The
pulse profile shows the clear double-peaked structure characteristic
of emission from both poles on the neutron star.  In addition, the
profile in gamma-rays closely resembles that exhibited by SMC X-1 in
the hard X-rays (Naik \& Paul 2004, Wojdowski et al 1998).

\begin{figure}
\begin{center}
\includegraphics[width=80mm,angle=-0]{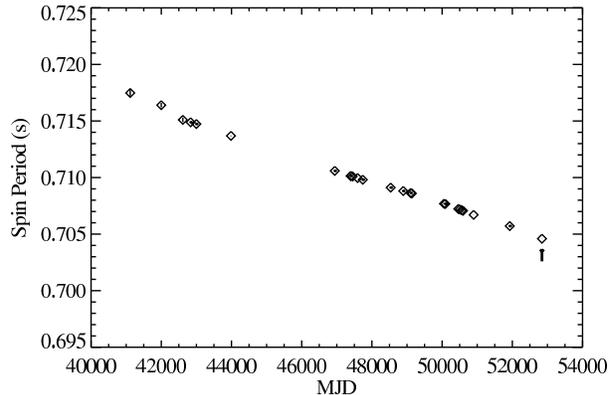}
\caption{Spin up of SMC X-1 over the last 30 years. The INTEGRAL
result is the last in the series and indicated by an arrow.}
\label{fig:spinup}
\end{center}
\end{figure}

SMC~X-1 has shown a constant spin up over the last thirty years of
observations.  Spin-up is believed to be an effect of accretion
torques (Ghosh \& Lamb 1979), i.e. matter orbiting in the accretion
disc around the neutron star transfers angular momentum to the neutron
star when it accretes, spinning up the neutron star.  Although some
accreting X-ray pulsars show periods of spin-up and spin-down, SMC~X-1
is persistently accreting matter from the stellar wind of its
supergiant companion hence it has been spinning up for at least as
long it has been observed.  Figure~\ref{fig:spinup} shows the evolution
of the spin period of SMC~X-1, with the arrow denoting the measurement
made in this work.

\subsection{Source 2}
\label{SectTimingSource2}

Apart from \hbox{SMC~X-1}, the INTEGRAL map reveals a second source which we
have designated here as Source 2.  In the Third IBIS/ISGRI Soft
Gamma-ray Survey Catalog (Bird et al, 2007), Source 2 was identified
with the nearest ROSAT source: RX~J0053.8$-$7226 at $00^{\rm h}
53^{\rm m} 53.4^{\rm s}$, $-72^\circ 27' 01''$.  
This source (=1WGA~J0053.8-7226
=XTE~J0053-724), a known Be/X-ray binary in the SMC
(Haberl \& Pietsch, 2004), has been observed with ROSAT, RXTE
and ASCA and has a well-established spin period of 46.6\,s
(Corbet et al, 1998).  

\begin{figure}
\begin{center}
\includegraphics[width=80mm,angle=-0]{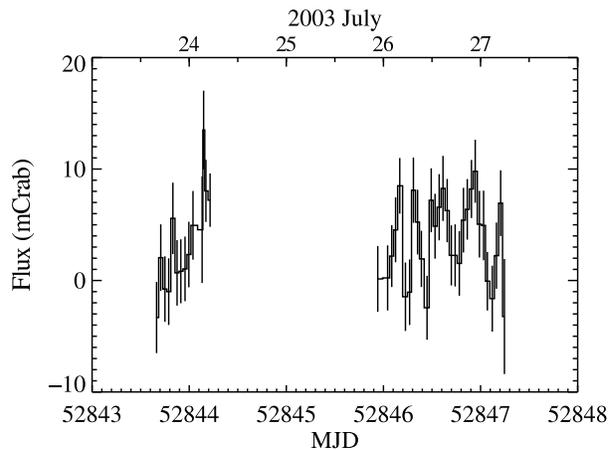}
\caption{Light curve of Source 2 at Science Window time resolution}
\label{fig:lc2}
\end{center}
\end{figure}

The long term lightcurve of Source 2 is shown in Figure~\ref{fig:lc2}
from MJD 52843.  During the first block of data (revolution 94) the
source appears to be in the process of switching on.

In order to search for the 46.6s spin period of \hbox{RX~J0053.8$-$7226}, we
extracted a lightcurve at 1\,s time resolution. Instead of creating
the source shadowgram, the lightcurves were assembled using the {\em
iilight} tool supplied with OSA 5.1.  This tool extracts a lightcurve
for every source as well as the background in the field of view by
using the Pixel Illumination Function (PIF).  The advantage of this
method is that lightcurves down to a time resolution 0.1\,s can be
constructed.  

\begin{figure}
\begin{center}
\includegraphics[width=60mm,angle=-90]{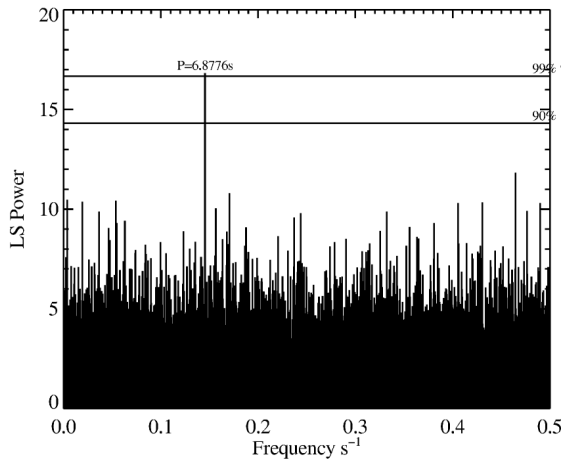}
\caption{Lomb-Scargle periodogram of Source 2. The solid lines show
the 90\% and 99\% confidence levels determined from the Monte Carlo
simulations.}
\label{fig:pdg2}
\end{center}
\end{figure}

The lightcurve was subjected to a Lomb-Scargle analysis to search for
periodicities in the range 2--1000\,s (Lomb 1976, Scargle 1982).
Although no significant peaks in the periodogram were detected during
revolution 94, or the combined lightcurve of revolutions 94 and 95, a
marginal periodicity at $6.878\pm 0.001$\,s was detected in revolution
95 (see Fig.~\ref{fig:pdg2}).  We also searched lightcurves binned at
2, 5 and 20\,s, but detected no periodicities above a significance of
90\%.

The significance levels overplotted in this Fig.~\ref{fig:pdg2} were
calculated from 10000 Monte Carlo simulations of the dataset.  Each
simulated lightcurve had the timestamps of the original lightcurve,
but data randomly drawn from a Gaussian distribution around the mean
of the original data and with the variance of the original data. A
Lomb-Scargle analysis was performed on each lightcurve, and the
maximum power for each of the 10000 lightcurves was noted, and used to
generate the cumulative distribution function.  From this distribution
one can read off the probability of finding a peak of given power in
the periodogram due to random noise. The 6.8s peak just exceeds the
99\% level (2.6$\sigma$).

To explore the significance of both the detection of a 6.8\,s period
and the non-detection of a 46.6\,s periodicity in the data, simulated
sine waves of differing periods were added to the existing lightcurve
such that the pulse fraction was 100\% and then put through the same
Lomb-Scargle analysis.  These simulations showed that, even at a
pulsed fraction of 100\%, a 46.6\,s periodicity would be unlikely to
show up above the noise.  For this same reason we note that the
detection of the 6.8\,s periodicity is only at the 99\% significance
level and may be spurious.


The non-detection of a 46.6s periodicity in the lightcurve from
Source~2 means that the identity of this source still remains an open
question.  Although it may yet be shown that Source~2 can be
identified with \hbox{RX~J0053.8$-$7226}, it is also a likely
possibility that Source~2 may be another X-ray binary in the SMC with
a pulse period $\sim6.8$\,s.  We explore these options further in
Sect.~\ref{SectDiscussion}.


\section{Spectral modelling}

\subsection{SMC X--1}

\begin{figure*}
\begin{center}
\includegraphics[width=100mm,angle=90]{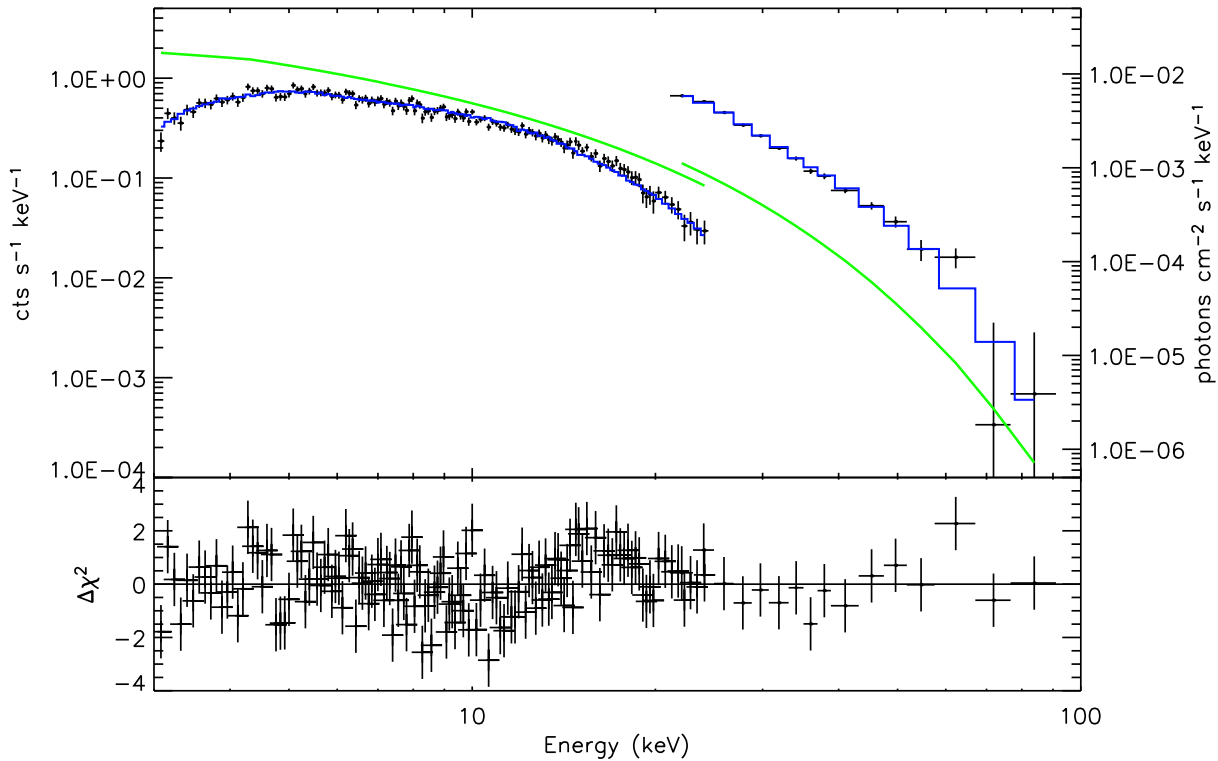}
\caption{3-100\,keV spectrum of SMC~X-1 fitted with a Comptonisation
model. The black crosses show the data points, 
the histogram the fitted model, and the grey line 
shows the unfolded model and is plotted on the secondary Y-axis.}
\label{fig:spect}
\end{center}
\end{figure*}

A source spectrum in the energy range 3--100\,keV is shown in
Figure~\ref{fig:spect}, using data from the JEM-X (3--25\,keV) and
IBIS (20--100\,keV) detectors.  This spectrum combines all spectra
from individual observations of SMC~X-1 during revolutions 94 and 95.
The JEM-X spectrum is generated by combining spectra from individual
science windows during which the source was within 7.5$^\circ$ of the
pointing axis.  During all these observations SMC~X-1 was in the high
state of its 40--60\,d super-orbital cycle (Clarkson et al., 2003), 
so spectral parameters
will vary from observations taken during the low state.  Spectral
fitting was performed using XSPEC v11.3.1 (Arnaud 1996).

We fitted thermal bremsstrahlung and power law models to the data.
Because the data do not cover energies below 3\,keV, in both cases we
constrained the absorption column between a lower limit of
$4.64\times10^{20}$atoms cm$^{-2}$, the estimated neutral hydrogen
density in a cone of radius 1$^\circ$ in the direction of the SMC
(Dickey \& Lockman 1990) and an upper limit of $9\times 10^{21}$\,atoms
cm$^{-2}$, the highest value of $N_{\rm H}$ recorded by
Woo et al., (1995).  The best fit ($\chi^2_\nu=1.14$) was
achieved using a photoelectric absorption component, with cross-sections from Balucinska-Church \& McCammon (1992) and abundances from Anders \& Grevesse (1989), with a high energy
cutoff power law model.  No emission from FeK$\alpha$, which is often
seen in X-ray binaries and has been noted in SMC~X-1 on a previous
occasion (Naik \& Paul 2004), was detected.  The spectral parameters
for the best fit model are given in
Table~\ref{TabSmcx1Specpars}.

\begin{table}
\caption{Spectral parameters for an exponentially cutoff power law model fit to SMC~X-1.}
\begin{tabular}[t]{ll|ll}
\hline
\noalign{\smallskip}
 \multicolumn{2}{c}{Our work} &
\multicolumn{2}{c}{Naik \& Paul (2004)}\\ 
Param & Value &Param & Value\\
\hline
\noalign{\smallskip}
$N_{\rm H}$ & $3.8^{+5.1}_{-3.3}\times10^{21}$cm$^{-2}$ 
& $N_{\rm H}$ & $2.55\pm0.09\times10^{21}$cm$^{-2}$\\
$\Gamma$ & $0.57\pm0.07$ 
& $\Gamma$ & $0.82\pm0.02$\\
$E_{\rm cut}$ & $4.3\pm0.5$\,keV 
& $E_{\rm cut}$ & $6.3\pm0.09$\,keV\\
$E_{\rm fold}$ & $9.2^{+0.1}_{-0.2}$\,keV
& $E_{\rm fold}$ & $10.58\pm0.13$\,keV\\
& & $kT$ & $0.19\pm0.01$\,keV\\
$\chi^2$(dof) & 168(145) 
& $\chi^2_{\nu}$(dof) & 1.3(168)\\
\noalign{\smallskip}
\hline
\end{tabular}
\label{TabSmcx1Specpars}
\end{table}

Comparing the ISGRI spectrum to previous studies of the broadband
spectrum of SMC~X-1 will allow us to notice whether there are any long
term spectral changes in the source.  Prior to these data, the most
recent broadband (0.1--80\,keV) spectral analysis of SMC~X-1 was
undertaken by Naik \& Paul (2004) using data obtained between 1997
January to April from the three detectors on board {\em BeppoSAX}.  A
best fit was obtained by modelling the data with a power law with a
high energy cutoff, a thermal blackbody component to account for
emission in the soft part of the spectrum, and a weak Fe line.

The data from the current observation extend down to 3\,keV, i.e. not
low enough in energy to allow us to investigate the soft blackbody
component that makes its major contribution below energies of
1.0\,keV.  Similarly, the column density for our observations is not
well constrained without data below 3\,keV.  Although the parameters
for the power law slope ($\Gamma$), and the $E_{\rm cut}$ differ
substantially from those obtained by Naik \& Paul (2004), if we fix
the $E_{\rm cut}$ at 6.3\,keV we can recover, within errors, the same
power law slope as measured in their observations.  Thus, in general,
our continuum models agree with those of Naik \& Paul (2004) and
hence we note that no significant changes in the continuum X- and
$\gamma$-ray emission took place between these two observations.

Naik \& Paul (2004), however, report the detection 
of a weak FeK$\alpha$ emission line, which we find no evidence for in
this observation.  Fe emission is also reported by
Woo et al. (1995) in a {\em Ginga} observation from 1989.
We attribute the non-detection of an Fe emission component to the
lower sensitivity of the JEM-X instrument.  The Fe line flux as
reported by Naik \& Paul (2004) is
$1.4\times10^{12}$\,erg\,cm$^{-2}$\,s$^{-1}$, which converts into
$1.5\times10^{-4}$\,photons\,cm$^{-2}$\,s$^{-1}$ at 6.4\,keV.  This is
below the JEM-X line sensitivity, which is
$1.6\times10^{-4}$\,photons\,cm$^{-2}$\,s$^{-1}$ at 6\,keV.  We note
that the lack of the Fe emission line at $\sim 6.4$\,keV in our
dataset may contribute to shifting the $E_{\rm cut}$ to softer
energies.

Kunz et al. (1993) examine the X-ray spectral 
properties of SMC~X-1 in the 20--80\,keV energy band during
observations by {\em HEXE} between 1987 November and 1989 March.  They
describe the continuum by fitting a thin thermal bremsstrahlung model.
Although our data are not fit at all well by the above model, if we
restrict our energy range to 20--80\,keV, using only the IBIS
spectrum, a thermal bremsstrahlung model gives an adequate fit and
plasma temperature of $12.2\pm0.4$\,keV; close to the plasma
temperature of $14.4\pm0.13$\,keV by Kunz et al. (1993).
We can conclude that taking the low energy data into account rules out
the thermal bremsstrahlung model.

\subsection{Source 2}

The ISGRI spectrum of Source 2 from 20--100\,keV was
best fit with a power law of slope 2.6 ($\chi^2$(dof)=7.2(3)).  Other
models, such as a power law with a high energy exponential cutoff,
Comptonisation, and thermal bremsstrahlung models were used
with less successful results.

Although a soft power law is the best model to describe the data in
the IBIS energy range, it may not be appropriate across the entire
energy range.  Measures of the absorption column and blackbody
contributions to the spectrum often come into play at energies between
0.1 and 3\,keV.  The source is not detected in $\sim 149$\,ks of
simultaneous JEM-X observations.  Using the power law parameters
derived from the fit to the IBIS data we estimate the flux in the
JEM-X 5--10\,keV range to be 
$\sim6\times10^{-11}$\,erg\,cm$^{-2}$\,s$^{-1}$. With a very bright source
(SMC~X-1) in the field of view it is clear that Source~2 is at or
below the JEM-X detection limit.

\section{Discussion}
\label{SectDiscussion} 
In this section we discuss the possible counterparts for Source~2.

A pulsar, XTE~J0103$-$728 = SXP6.85, with spin period $6.8482 \pm
0.0007$\,s has previously been reported in the SMC (Corbet et al.,
2003).  On the basis of the similarity of their periods it is tempting
to assume that this object is the same as the INTEGRAL
Source~2. However, recent XMM-Newton observations (Haberl et al. 2007)
have located this pulsar to within $2\arcsec$ at a position some
$40\arcmin$ away from the INTEGRAL source (see
Fig.~\ref{fig:xtej2a}). Furthermore, Galache et al. (2007) report, in
their RXTE monitoring of pulsars in the SMC, that SXP6.85 was not
detected around the time of the INTEGRAL observations.  The nearest
RXTE measurement (about 30--40\,d later) confirms that the periods are
close, but distinctly different.  Based on the clear discrepancies in
both the positions and the pulse periods, we conclude that
XTE~J0103$-$728 can not be the counterpart to Source~2.


\begin{figure}
\begin{center}
\includegraphics[width=90mm,angle=0]{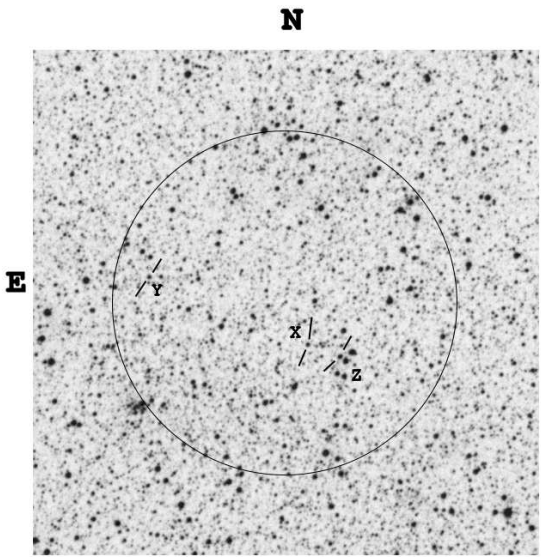}
\caption{ Finding chart showing the position of the INTEGRAL 90\%
uncertainty circle for Source 2. The circle has a radius of 3.4
arcmin. Indicated as X, Y \& Z are the three bright (V$\sim$15) 
optical objects that coincide with Chandra X-ray sources.}
\label{fig:fc2}
\end{center}
\end{figure}

Chandra X-ray observations of the region around Source~2 during 2002
July and 2006 April detect 11 X-ray sources within the INTEGRAL error
circle, of which three correlate with previously known
objects in the Simbad database and have V$\sim15$ optical
counterparts.  These objects are labelled X, Y and Z in
Figure~\ref{fig:fc2} and have the following identifications:
\begin{itemize}
\item[(i)]Object X at $00^{\rm h}54^{\rm m} 03.7^{\rm s},-72^\circ 26'33.1''$ is 
SMC 28942 from Massey (2002) with $V=14.81$, $B-V=-0.04$ 

\item[(ii)]Object Y at $00^{\rm h}54^{\rm m} 46.2^{\rm s},-72^\circ 25'23.0''$ is SMC 31155 from Massey (2002) and identified as object 798 in the catalogue of Meyssonnier \& Azzopardi (1993) with $V=15.25$, $B-V=-0.10$. 

\item[(iii)]Object Z at $00^{\rm h}53^{\rm m} 55.2^{\rm s},-72^\circ 26'45.7''$ is SMC 28479 from Massey (2002) with $V=13.98$, $B-V=-0.09$.
\end{itemize}

All sources were too faint in the Chandra X-ray data to perform a
timing analysis, which would have been the most direct means of
linking any of them with Source~2.

\begin{figure}
\begin{center}
\includegraphics[width=80mm,angle=0]{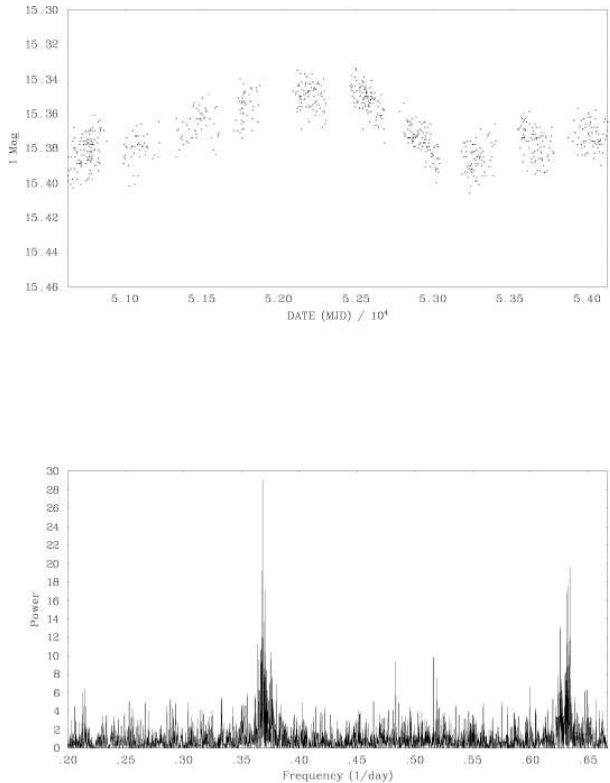}
\caption{Top: the OGLE I-band light curve over 10 years for Object
Y. Lower: the associated Lomb-Scargle power spectrum of Object Y
showing a peak at a period of $\sim$2.7\,d.}
\label{fig:pdp}
\end{center}
\end{figure}

Long-term OGLE optical photometric data in the $I$ band were examined
for all three objects. Objects X \& Z exhibit significant variability
patterns of 0.3--0.5 mag over several years, characteristic of the
behaviour of Be stars (Mennickent et al., 2002). However, subjecting
the data to a frequency search using a Lomb-Scargle routine reveals
nothing periodic in the search range of 1.5 -- 200 days.  The third
object (Y), however, shows much smaller changes of the order of 0.05
mags over many years (see Figure~\ref{fig:pdp}). More importantly, the
Lomb-Scargle search reveals a strong and clear modulation at a period
of 2.71443$\pm$0.00012\,d.  Also evident in Figure~\ref{fig:pdp},
which shows the results from searching the 
period range 1.5--5\,d, is the beat period of the
2.71\,d with the 1\,d sampling.

Chandra source Z has been firmly associated with the Be/X-ray binary
system \hbox{RX~J0053.8$-$7226} =\hbox{1WGA~J0053.8$-$7226}
=\hbox{SXP46.6} (Coe et al. 2007), which has a pulse period of
46.6\,s.  As can be seen from Fig~\ref{fig:fc2}, the source position
is consistent with the IBIS position of Source~2.  Futhermore, RXTE
has observed \hbox{RX~J0053.8$-$7226} in outburst during a Type~I
periastron passage at the time of the INTEGRAL observation (Galache et
al. 2007).  Extrapolation of the X-ray flux from
\hbox{RX~J0053.8$-$7226} observed during this outburst into the IBIS
20--40\,keV energy range (estimating a pulsed fraction of 0.2 and a
photon index of 1) shows that \hbox{RX~J0053.8$-$7226} should be
detected at the $11\sigma$ level with IBIS -- i.e. consistent with the
current level at which Source~2 has been detected.

As the spatial and flux information, as well as the epoch of
observation of Source~2, are all compatible with reasonable
assumptions about the behaviour of \hbox{RX~J0053.8$-$7226}, the
simplest conclusion is that Source~2 as observed by IBIS can be
identified with \hbox{RX~J0053.8$-$7226} as initially suggested by
Bird et al (2007).  However, this conclusion is by no means secure, as
equally reasonable extrapolations of the RXTE flux from
\hbox{RX~J0053.8$-$7226} (using a pulsed fraction of 0.5 and a photon
index of 2) yield a source that would be well below the detection
limit of IBIS.  In addition to this, it is nescessary to account for
the lack of detection of a 46.6\,s periodicity, and instead the
possible detection of a 6.8\,s periodicity in the data from Source~2.

The simplest explanation is that a low pulse fraction, combined with
the low flux source detected by IBIS, makes the 46.6\, pulsations
undetectable.  This does not, however, explain the presence of the
6.8\,s pulsations detected in the IBIS data.  If this detection is
real then \hbox{RX~J0053.8$-$7226} cannot be the counterpart to
INTEGRAL Source~2.  The simulations we have performed in
Sect.~\ref{SectTimingSource2} indicate that the detection of
pulsations is marginal at this flux/exposure, and requires an
extremely high pulsed fraction for this particular dataset.  While the
power of the 6.8\,s pulsations is strong, we cannot rule out the
possibility that the detection is spurious.

With these concerns raised, we also consider possible scenarios in the
case that \hbox{RX~J0053.8$-$7226} (Chandra source Z) is not the
counterpart to Source~2.

We have little information on Chandra source X except that its optical
properties resemble those of source Z.  We could consider this a point
in favour of X being the counterpart to IBIS Source~2, which may then
indicate that Source~2 is likely to be a Be/X-ray binary.

\begin{figure}
\begin{center}
\includegraphics[width=70mm,angle=-90]{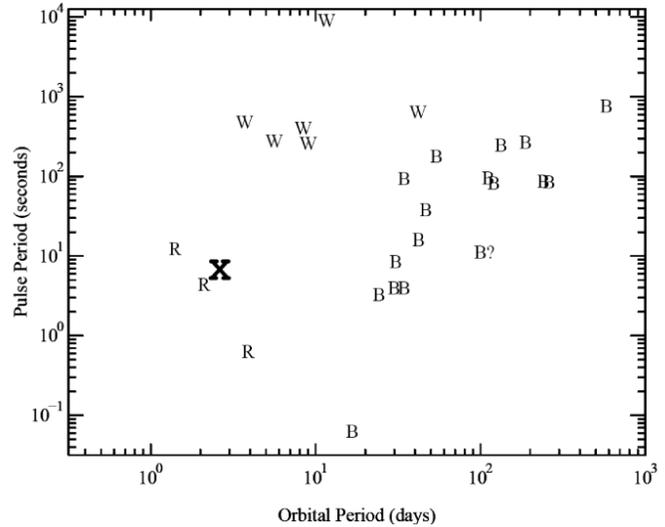}
\caption{The Corbet diagram (modified from Corbet et al. 1999) 
showing the location
of INTEGRAL Source 2 marked with the X symbol (see text for
assumptions made in locating the source at this position). 
The B symbol indicates
a Be star system, the W indicates a wind-fed system and the R symbol
indicates the Roche-lobe overflow systems.}
\label{fig:corbet}
\end{center}
\end{figure}

If Chandra source Y, with its 2.7\,d optical periodicity were the
counterpart to Source~2 (with its 6.8\,s pulse period), this object
would occupy a unique corner of the Corbet diagram (see
Fig.~\ref{fig:corbet}), indicating that the source may be a Roche Lobe
Overflow (RLOF) system.  Thus far only three RLOF systems are known:
SMC~X-1, Cen~X-3 and LMC~X-4 and adding another object to this class
of sources would be an exciting prospect.  To investigate this
possibility we examine some of the X-ray and optical properties of
Chandra source Z and IBIS Source~2 and compare them with those of the
known RLOF systems.

A search through the Chandra archives reveals three observations of
this region on 2002 July 20, 2006 April 25 and 26.  Chandra source Y
was detected on all three occasions and exhibits a flux varying by a
factor of $\sim10$.  Existing RLOF systems display persistent emission
which varies between high and low states at superorbital periods.  If
Source~2 and Chandra source Y are indeed the same source, they have
been observed, at varying flux levels, in all four observations of this
region.

The optical magnitudes and colours of source Y are consistent with
those of an early-type, luminosity class III-V star, similar to the
optical counterpart of LMC~X-4 (Hutchings, Crampton \& Cowley 1978).
Furthermore, when X-ray colours are used to locate this object on the
colour-colour diagram for SMC sources (McGowan et al, 2007), source Y
sits comfortably in the middle of the known pulsar systems.

\section{Conclusions}

This work has shown that SMC~X-1 continues its secular spin-up trend,
with the latest pulse period measurement being
$0.704596\pm0.000008$\,s.  We can also conclude that the
broadband spectrum of SMC~X-1 is roughly consistent from 1987 through
2004.

We have discovered a potential 6.8\,s periodicity in the IBIS data
from Source~2, which casts some doubt on its current identification
as the counterpart to \hbox{RXJ~0053.8-7226}.  We conclude on the
basis of spatial and timing information that Source~2 is not related
to \hbox{XTE~J0103$-$728} -- a previously known Be/X-ray binary with a
pulse period of 6.85\,s. Archival Chandra observations are used to
investigate the feasibility of other X-ray/optical counterparts to the
sources.  We can conclude from this work that it is highly probable
that Source~2 is a HMXB and, depending on its X-ray/optical
associations, may yet turn out to be one of a small number of Roche
Lobe Overflow HMXB systems.  If this were the case, it is intriguing
both why Source~2 has not been detected with previous X-ray missions
(could it be obscured somehow in the softer X-rays?) and also, why
three out of the four known RLOF systems have evolved in the
Magellanic Clouds.

To resolve this situation, detection of either the 46.6\,s period in IBIS Source~2, or the 6.8\,s period in either of the Chandra sources X or Y would be required and further observations are encouraged.

\section{Acknowledgements}

VAM acknowledges support from the South African NRF and the British
Council in the form of a SALT/Stobie studentship.


\begin{thebibliography}{}

\bibitem[]{}Anders E. \& Grevesse N. 1989, Geochim. Cosmochim. Acta, 53, 197

\bibitem[]{}Arnaud K.~A. 1996,ASP Conf. Ser. 101: Astronomical Data Analysis Software and Systems V, eds. Jacoby G.H. and Barnes J.

\bibitem[]{}Balucinska-Church M. \& McCammon D. 1992, ApJ, 400, 699 

\bibitem[] {} {Bird} A.~J. et al.,  2007, ApJS 170, 175.


\bibitem[]{} Clarkson W.I., Charles P.A., Coe M.J., Laycock S., Tout
M.D. and Wilson C.A., 2003 MNRAS 339, 447.

\bibitem[] {} Coe, M. J., Edge, W. R. T., Galache, J. L. \& McBride,
V. A. 2005 MNRAS 356, 502.

\bibitem[] {} Coe, M.J., McGowan, K.E. and Schurch, M.P.E., 2007, in preparation.


\bibitem[] {} Corbet, R. H. D.; Marshall, F. E.; Peele, A. G.;
Takeshima, T. 1999 ApJ 517, 956.

\bibitem[] {}{Corbet} R.,  {Marshall} F.~E.,  {Lochner} J.~C.,  {Ozaki}
M.,  {Ueda} Y., 1998, IAU Circular, 6803, 1

\bibitem[] {}{Corbet} R.~H.~D.,  {Markwardt} C.~B.,  {Marshall} F.~E.,  {Coe} M.~J.,  {Edge}  W.~R.~T.,    {Laycock} S.,  2003, The Astronomer's Telegram, 163, 1

\bibitem[]{} {Dickey} J.~M.,  {Lockman} F.~J.,  1990, ARA\&A, 28, 215.

\bibitem[]{} Dolphin, A. E., Walker, A. R., Hodge, P. W., Mateo, M.,
Olszewski, E. W.. Schommer, R. A., Suntzeff, N. B. 2001 ApJ 562, 303.

\bibitem[] {} Galache J.L., Corbet R.H.D., Coe M.J., Schurch M. \&
Laycock S.G.T. 2007, ApJ (submitted).

\bibitem[]{} {Ghosh} P.,  {Lamb} F.~K.,  1979, ApJ, 234, 296

\bibitem[]{} Goldwurm, A., {David}, P. and {Foschini}, L. et al. 2003, A\&A 411, 223

\bibitem[]{} Haberl F. \& Sasaki M., 2000, A\&A 359, 573.

\bibitem[]{} {Haberl} F.,  {Pietsch} W.,  2004, A\&A, 414, 667

\bibitem[]{} Haberl F., Pietsch W. \& Kahabka P., 2007, ATEL 1095.

\bibitem[]{} Hutchings J.B., Crampton D. \& Cowley A.P., 2007, ApJ 225,548

\bibitem[] {}{Kunz} M.,  {Gruber} D.~E.,  {Kendziorra} E.,  {Kretschmar} P.,  {Maisack} M.,  {Mony} B.,  {Staubert} R.,  {Dobereiner} S.,  {Englhauser} J.,  {Pietsch} W.,   {Reppin} C.,  {Trumper} J.,  {Efremov} V.,  {Kaniovsky} S.,  {Kusnetzov} A.,     {Sunyaev} R.,  1993, A\&A, 268, 116

\bibitem[] {}{Leahy} D.~A.,  {Darbro} W.,  {Elsner} R.~F.,  {Weisskopf} M.~C.,  {Kahn} S.,  {Sutherland} P.~G.,    {Grindlay} J.~E.,  1983, ApJ, 266, 160

\bibitem[]{} {Lomb} N.~R.,  1976, Ap\&SS, 39, 447

\bibitem[] {} Massey P., 2002 ApJS 141, 81

\bibitem[] {} McGowan K.E. et al., 2007 MNRAS 376, 759.

\bibitem[]{} Mennickent R.E., Pietrzynski G., Gieren W. \& Szewczyk
O., 2002 A\&A, 393, 887.

\bibitem[]{} Meyssonnier N. \& Azzopardi M., 1993 A\&AS 102, 451

\bibitem[]{} {Naik} S.,  {Paul} B.,  2004, A\&A, 418, 655


\bibitem[]{} {Scargle} J.~D.,  1982, ApJ, 263, 835

\bibitem[]{} {Wojdowski} P.,  {Clark} G.~W.,  {Levine} A.~M.,  {Woo} J.~W.,    {Zhang}  S.~N.,  1998, ApJ, 502, 253

\bibitem {} Woo J.~W.,  {Clark} G.~W.,  {Blondin} J.~M.,  {Kallman} T.~R.,    {Nagase}  F.,  1995, ApJ, 445, 896


\end{thebibliography}
\end{document}